\documentstyle[11pt]{article}
\title{Some comments on\\
Models of hadronic interactions at air shower energies}
\author{T.K. Gaisser\\
Bartol Research Institute, University of Delaware, Newark, DE 19716}

\begin{document}
\maketitle
\begin{abstract}
Several models of minimum-bias hadronic interactions at ultra-high energy
that have been used for calculations of air showers share essential
common features.  In this talk I review these common elements and
discuss some consequences.  I concentrate on
properties of hadron-nucleus interactions, and I use
mean depth of shower maximum as a function of primary energy
to illustrate my main points.  I will contrast these models
with models that use a more naive treatment of hadronic
interactions in nuclei but which have been successfully
used to interpret measurements of depth of shower maximum.
\end{abstract}
\section{Astrophysical motivation}
The spectrum of cosmic rays  extends more than five orders of magnitude
beyond the highest energy at which it has been possible so far
to observe the primaries directly with experiments on balloons
or spacecraft.  In this energy range the spectrum has at least
two features, the knee between $10^{15}$ and $10^{16}$~eV and the
ankle between $10^{18}$ and $10^{19}$~eV.  The knee may be associated
with a transition between different classes of galactic cosmic
rays (or with a feature of propagation of galactic
cosmic rays) and the ankle with a transition from galactic to
extra-galactic origin of the particles---but these are only
plausible conjectures.

In order to understand
the implications of these features for the origin of the high-energy
particles it is necessary to measure the relative contribution
of the different groups of nuclei.  This must be done with large
ground-based air-shower experiments in order to achieve sufficient exposure
to  collect a large sample of high-energy events.
Because of the indirect nature
of air shower experiments and the problems of fluctuations
superimposed on a steeply falling energy spectrum,
progress toward the goal of measuring the primary composition
at high energy has been slow and difficult.
New experiments with better resolution, coupled with a
better understanding of the hadronic interaction models
used to interpret the data, show promise for improving
the situation.

\subsection{The knee region}

The current status is summarized very nicely in the
review of Kalmykov and Khristiansen \cite{KK}.
Concerning the knee region, they conclude that measurements
of fluctuations in the muon to electron ratio suggest
a gradual change toward a larger fraction of heavy nuclei
in the energy range from $10^{15}$ to $10^{17}$~eV.
They consider two quite different models which share
this feature.  One is a diffusion model \cite{Ptuskin}
in which a single underlying composition and source
spectrum are modified by propagation in the galactic magnetic field.
The other is a model \cite{Biermann,Stanevetal,Biermannetal} in
which there are two different classes of sources.

The diffusion model is a modern realization of the classic
suggestion of Peters \cite{Peters} and Zatsepin \cite{Zatsepin}.
Its characteristic feature is a steepening of the spectrum
at a certain value of magnetic rigidity so that there
are relatively more heavy particles at high energy
when the classification is by total energy per nucleus.
To fit the data the break has to occur at a rigidity
of $\ge 10^{6}$~GV.

The two-component model is motivated by the observation \cite{Swordy}
that the proton spectrum as measured by JACEE \cite{JACEE}
appears to steepen around 100 TeV or
somewhat below while there is no steepening of helium or
heavier nuclei at the corresponding rigidity.  The models are
further stimulated by the fact that 100 TeV is the maximum
energy expected for shock acceleration by supernova blast waves
expanding into the typical interstellar medium \cite{LagageCesarsky}.
In this kind of model higher energy particles come from acceleration
by supernovae whose environments are richer in heavy nuclei,
so the high-energy component contains fewer protons and relatively
more heavy nuclei.  The specific version of this model
described in Ref. \cite{Stanevetal} has less than 10\% protons
already for $E>300$~TeV.  According to Kalmykov and
Khristiansen the muon/electron measurements rule out such a
low fraction of protons below the knee, presumably because
without the light nuclei one would see smaller fluctuations
in the muon/electron ratio than observed.

Most likely, the real situation combines features of both
diffusion and different types of sources.  In addition,
one would expect supernovae in the same general class to
have somewhat different environments (magnetic fields
and composition) and thus to exhibit a gradation of
cutoffs and compositions of the cosmic-ray spectra they
produce.  Nevertheless it is
interesting to look at the mass fractions in these specific models just
to illustrate how well experiments will need to determine the mass
composition in order to distinguish among models.
If the composition is 64\% protons and helium and 36\%
heavier nuclei below the knee (``normal composition'')
and all components have a differential spectral index of -2.7,
then the corresponding fractions above the knee (e.g.
around $10^{17}$~eV) would be 44\% and 56\% if the spectral
index changes by -0.3 to -3.0.  The corresponding numbers
given in Ref. \cite{KK} for the Hall diffusion model of
Ptuskin et al. \cite{Ptuskin} are 55/45 below the knee and
28/72 at $10^{17}$~eV.  The light/heavy ratio is similar
to this at high energy in the two-source model of Ref. \cite{Stanevetal},
but the light component is reduced before the knee as well.

\subsection{Energies above $10^{17}$~eV}

An analysis of the Fly's Eye data \cite{Bird} on mean
depth of shower maximum as a function of primary energy
leads to the conclusion that the primaries are mostly
heavy nuclei at the low end of the energy range of this
experiment ($10^{17}$~eV), with an increasing fraction of protons
as energy increases.  It was noted \cite{Bird} that
this transition corresponds in energy with the ``ankle''
feature of the spectrum.  If it is correct, this interpretation is
circumstantial evidence for a transition to extragalactic
origin for the highest energy cosmic rays, and this
extragalactic component is primarily light nuclei
(protons and helium).

In their review, Kalmykov and Khristiansen reach a similar
conclusion, although the evidence for a change of compositon
is less strong than indicated by the figure of Ref. \cite{Bird}.
Yakutsk data are consistent with those of the Fly's Eye group
\cite{Yakutsk,KK}.  In fact various calculations give a wide
range of results for depth of maximum in the Fly's Eye energy
range.  After discussing some of the input to the models,
I will return to this question in the last section.

\section{Interaction models for EAS}

Comparison to Fly's Eye data, with typical energies of
$10^{18}$~eV and above, requires extrapolation of hadronic
interaction models more than two orders of magnitude
in center of mass energy beyond the highest accelerator
energies ($\sqrt{s}=1.8$~TeV).  In fact the required extrapolation
is much greater than this because the showers involve
nuclei as well as single hadrons (both as targets and as
projectiles) and because showers depend essentially on
how the energy is divided when particles interact (fragmentation
region) rather than on multiplicity (central region).  Physical
models are required to relate the various processes and to
extrapolate to high energy.  Thus also in the knee region
there can be significant model-dependence in air shower simulations.

In preparation for this meeting Knapp et al. \cite{Knapp} have made a very
instructive comparison of some results of different
event generators, several of which represent theoretical
models of hadronic interactions with similar underlying physical
assumptions.  The models (codes) that I consider here are:

\begin{enumerate}
\item Dual parton model \cite{cap} (DPMJET \cite{Ranft}).
\item Quark-Gluon String model \cite{QGS} (QGSjet \cite{QGSjet}).
\item Venus \cite{Werner}.
\item Minijet model \cite{Durand,GS} (SIBYLL \cite{SIBYLL}).
\end{enumerate}
All of these codes use strings of hadrons coupled to constituent
partons of the incident hadrons to represent particle production
in high-energy interactions.

\subsection{Inelasticity}

The common feature of all these models
that I want to draw attention to concerns
interaction of a single hadron with a target nucleus and, in particular,
the propagation of the projectile hadron and its fragments
through the target nucleus.  In all the models listed above
the valence partons in the projectile only radiate one pair of strings
of secondary hadrons.  In cases with more than one wounded nucleon
in the target the extra strings are connected with sea-quarks in the
projectile.  This means that the inelasticity in hadron-nucleus
collisions is not much larger than that in corresponding hadron-nucleon
collisions.  This is illustrated by comparison of Figs. 14 and 35
in the report of Knapp et al. \cite{Knapp}.

One can contrast this, for example, with the KNP model \cite{KNP},
which was favored by the Fly's Eye group in its comparison \cite{Bird} of
measurements of depth of maximum between $10^{17}$ and $10^{19}$~eV
with simulations.  Cronin (private communication) has emphasized
that other codes, such as SIBYLL and MOCCA \cite{MOCCA}, give
a rather different impression of the interpretation of the
measurments than that shown in Fig. 2 of Ref. \cite{Bird}.
In that figure the Fly's Eye data are compared to mean depth of
maximum vs. energy for proton showers and for iron showers as
simulated with the KNP model.  This choice was made because, of
the several models compared with the same data in an earlier paper
\cite{Gaisser}, the KNP model gave the most rapid shower development
and hence the best agreement with the observed depth of maximum.

One feature of the KNP model that led to relatively rapid shower
development is the treatment of propagation of the projectile
nucleon inside the target nucleus.  In this model the projectile
nucleon is assumed to lose energy in the same way each time
it interacts with a target nucleon in the nucleus.  This leads
to collisions that are significantly more inelastic. 

Inelasticity is the fraction of energy in a hadron-nucleus
interaction {\em not} carried off by the fragment of the
projectile hadron. Because there is no way to distinguish
a ``fragment'' hadron from a ``produced'' hadron of the same type, this is
only a theoretical definition (but still a useful one for
comparison of theoretical models).
The following definition of inelasticity, $K$, in
high-energy proton-nucleus collisions in terms of
momentum-weighted integrals of inclusive cross sections
is one that could in principle be realized experimentally:
\begin{eqnarray}
1\,-\,K\;\approx\;&\left(\int\,p\,{d\sigma_N\over d^3p}d^3p\right . \\
&\;\;-\;\left.
\int\,p\,{d\sigma_{\bar{N}}\over d^3p}d^3p\right)/(\sigma\times E_0).\nonumber
\end{eqnarray}
Here $\sigma_N$ is the cross section for production of a proton
or neutron and $\sigma_{\bar{N}}$ the cross section for producing
an antinucleon.  The definition becomes more accurate as energy
increases and the target fragments become relatively less important.
An experimental definition is more difficult for pions, but what counts
is the inclusive cross section in the fragmentation region
(e.g. for $x>0.05$).

H\"{u}fner and Klar \cite{HK} compared a simple  model
with data on p-nucleus interactions in the 100 GeV region
\cite{Barton,deMarzo}.  Within the context of a Glauber
multiple scattering treatment of hadron-nucleus collisions
the mean number of wounded nucleons in the target is given by
\begin{equation}
\langle N_W\rangle_{pA}\;=\;{A\sigma_{pp}^{inel}\over\sigma_{pA}^{inel}},
\label{Nwound}
\end{equation}
and the probability, $P_n$ of having exactly $n$ wounded nucleons
is calculated from the nuclear profile functions \cite{Glauber}
in the standard
way \cite{SIBYLL}.  H\"{u}fner and Klar
write the energy of the projectile fragment nucleon after
encounters with exactly $n$ wounded nucleons on average as
\begin{equation}
E_N(n)\;=\;(1\,-\,I_{n})\times E_N(n-1).
\end{equation}
The total inelasticity is then
\begin{equation}
1\,-\,K\;\approx\;\sum_{n=1}^A P_n\prod_{k=1}^n (1-I_k).
\label{inelasticity}
\end{equation}
In this relation $I_1\approx {1\over 2}$ is the average inelasticity for
proton-proton collisions.  Within the string models listed
above one expects $I_1>I_2\approx I_3 \approx\ldots$.
Comparison of this simple formulation to the data in Ref. \cite{HK}
gives $I_n\approx 0.2$ for $n\ge 2$.
The KNP model has instead,$$
I_1 \;=\;I_2\;=\;I_3\;\ldots\approx 0.5
$$ and hence a larger value of inelasticity.

Figure 1 shows a plot of
inelasticity as a function of $\log(E_0)$ in eV
for the two sets of parameters calculated from Eq. \ref{inelasticity}.
The solid line corresponds to \{$I_n=0.5$ for all $n$\};
the dashed line to \{$I_1=0.5,\;I_n=0.2,\,n\ge 2$\}.
In both cases the inelasticity increases with energy.  This is
a consequence of the increasing cross section, which, according
Eq.~\ref{Nwound}, drives an increase in the number of wounded
nucleons.  The same cross sections \cite{SIBYLL} and hence
the same distributions of wounded nucleons have been
used for both curves in Fig. 1.  The difference is in the
assumption for $I_n,~n\ge 2$.  The upper curve represents the KNP
model, and the lower curve is a good approximation to the
inelasticity in the full SIBYLL model.

\subsection{SIBYLL model}

This model is based on the idea that the increase in
cross section is driven by the production of minijets
\cite{CHL,GH,PS}.  The emphasis is on the fragmentation
region and on collisions of hadrons with light nuclei.
Depth of shower maximum and TeV muons are
the principal intended applications.
With one significant
exception, all the ideas of the dual parton model are incorporated.
The program uses Lund techniques \cite{Sjostrand} and is tailored
for efficient operation up to at least $10^{20}$~eV.

The exception is that in SIBYLL the soft part of
the eikonal function is taken as a constant, fitted to
low energy data.  The non-jet part of every event is
represented by exchange of a single pair of strings.
In the other string-type models
listed above, non-jet contributions to events can involve exchange of
multiple pairs of soft strings.  To the extent
that the parameters in the two approaches are tuned to
reproduce the same increasing cross section, the resulting
multiplicity should be similar---in one case it comes from
short pairs of multiple soft strings and in the other from
minijets.  Since the fragmentation of the minijet pair is
treated as a loop of string, the multiplicity should be similar.
In principle, the soft part of the eikonal should also
be energy-dependent, and multiple exchanges of both hard
and soft kinds should occur.
The question of sensitivity of the parameters of the fits to
data when both multiple soft and multiple hard exchanges are allowed is
an interesting one.

  For SIBYLL
at a lab energy of 2 TeV we find \cite{Frichter}
$$
I_1\approx 0.53\;\;I_2\approx 0.22\;\;I_3\approx 0.15\;\;I_4\approx 0.13.
$$
The overall inelasticity from SIBYLL for proton-nitrogen collisions is
shown by the inverted triangles in Fig. 1.  It is very close to
the simple formula \ref{inelasticity} with the partial
inelasticities from Ref.~\cite{HK}.

Because the inelasticities tend to converge as energy
decreases, it is not clear with what certainty the comparison
of Ref. \cite{HK} to the data restricts the behavior of the
inelasticity and even whether the KNP treatment is completely
ruled out.  This question needs futher investigation.

\section{Muon to electron ratio in air showers}
Interpretation of  measurements of the muon to electron ratio in air showers
deep in the atmosphere depend on Monte Carlo simulations.
Differences in
existing codes at present may make it difficult to
use this feature alone to discriminate among light/heavy
ratios of the kind illustrated in \S 1.1 above (see for example
Fig. 71 of Ref. \cite{Knapp}.   Obvious remedies are to
understand and resolve differences among the simulation
codes and to measure more components of the showers, as is
being done by KASCADE.  In the work referred to in Ref. \cite{KK},
the model-dependence is addressed by considering fluctuations
in the muon to electron ratio as well as the mean values.

In comparing production of $\sim$~GeV muons in cascades simulated
with different interaction models it may be useful to
tabulate the distribution of interactions of various types
in the cascade.  This diagnostic relates the
energy-dependence of low energy muons in showers to fundamental
features of the interaction models.  Generally it will be
possible to express the distribution of hadronic interactions
in a proton-induced air shower by an expression of
the form \cite{TKG}
\begin{eqnarray}
{d N_{int}\over d z}\;\approx\;& \delta(1-z) + f_p(z) +
a_\pi{(1-z)^{n_\pi}\over z^{1+p_\pi}}\\
& + a_K{(1-z)^{n_K}\over z^{1+p_K}},\nonumber
\end{eqnarray}
where $z = E_{int}/E_0$,  $f_p(z) \sim 1/z$ and
$E_0$ is the primary energy.  The
first two terms on the right side of Eq. 1
represent the interactions of nucleons in
the cascade, and the approximate expression for $f_p$ is
exact for a flat inelasticity distribution and neglecting
production of $N\bar{N}$ pairs.  The 3rd and 4th terms
represent interactions of pions and kaons respectively.

An approximation for the number of low-energy muons
in a shower initiated by a proton when
$E_\mu\ll \epsilon_\pi$ is \cite{TKG}
\begin{equation}
N_\mu(>E_\mu)\;\sim\;a_\pi\,F_{\pi\pi}(0)\,\ln{\epsilon_\pi\over E_\mu}
\left({E_0\over\epsilon_\pi}\right)^{p_\pi}
\label{Nmu}
\end{equation}
(plus a similar term for kaons).  Here
$\epsilon_\pi\approx 115$~GeV is the critical energy
for pions to decay rather than interact in the atmosphere.
The quantity $F_{\pi\pi}$ is approximately the rapidity
density in the central region for production of charged pions
in interactions of pions with nuclei of the atmosphere evaluated
for pion interaction energy of $\approx\epsilon_\pi$.  In a
superposition approximation, the A-dependence of the muon
content of a shower will be
\begin{equation}
N_\mu \;\approx\;A^{1-p_\pi}
\label{NmuA}
\end{equation}
for $E_{total}/A\gg\epsilon_\pi$.  The accuracy of these
approximations needs to be checked with simulations, but
the relations should be a useful guide to investigating
the model-dependence of the ratio of low energy muons to electrons
in air showers.

The Akeno group have published an analysis of low-energy
muons in large showers at AGASA \cite{Akeno}.  They fit
the muon  dependence on primary energy to a power
$$\langle \rho_\mu(600)\rangle\;\sim (S_{600})^p$$
with $p\approx 0.82$.  Here $S_{600}$ is the scintillator density
at $600$~m from the shower core,
which is assumed to be a good measure of the primary energy,
and $\rho_\mu$ is the corresponding muon density.
A change in composition of the
degree suggested by Fly's Eye would require a value of $p$
larger by roughly 10\%.  (The analysis involves a
comparison of Eqs. \ref{Nmu} and \ref{NmuA} above.)
Because of possible model-dependence of the parameters $p_\pi$
(and $p_K$) there is a corresponding model-dependence of
of the $\mu/e$ ratio which needs further investigation.

\section{Model-dependence of depth of maximum}
In this section I return to the question of the interpretation
of the depth of shower maximum as measured by Fly's Eye.
Figure 2 shows the Fly's Eye data \cite{Bird} and comparisons
to various model calculations.
The upper three lines in Fig. 2 are for
proton-induced showers calculated in three models:
SIBYLL---dotted \cite{SIBYLL}; QGS---dashed \cite{QGS};
KNP---highest solid line.  The representation of the
KNP model \cite{KNP} used for the Fly's Eye calculation
is described in Refs. \cite{Gaisser,Serap}.
Comparison of these three curves gives an indication of
the extent of the uncertainty in interpretation of
the depth of maximum measurements arising from uncertainties
in the models, such as different inelasticities and different
cross sections and interaction lengths.

In addition,
an important point is that in the Fly's Eye analysis of
Bird et al. \cite{Bird}
the simulations have been corrected for effects of detector acceptance
and other systematics.  The lower pair of solid lines in
Fig. 2 (taken from Ref.~\cite{Bird}) 
shows the KNP model for protons and for iron primaries
{\em after} these corrections have been applied.
There are several compensating effects, but the net shift
of about $-20$ to $-25$ g/cm$^2$ is shown by comparison to
the unshifted KNP curve for protons.

The recent work of the Utah group \cite{Ding} represents
a quite different approach to the calculation of depth of shower maximum.
They have analyzed the data in the context of a Chou-Yang
model \cite{Chou-Yang}.  This interaction model features
a statistical picture of particle production in the central
region according to which the inclusive cross section is
of the form
\begin{equation}
{d^3n\over d^3p}\;=\;{C\over E}\,e^{-\alpha p_T}\,e^{-E/T_p}.
\label{ChouYang}
\end{equation}
The parameters $C$ (normalization), $2/\alpha$ (mean transverse
momentum) and $T_p$ (partition temperature) are adjusted
to fit collider data in the central region from $\sqrt{s}=53$~GeV
to $1800$~GeV and extrapolated to higher energies
as described in Ref. \cite{Ding}.  The mean inelasticity
is the integral of Eq.~\ref{ChouYang}, which is approximately
\begin{equation}
K\;=\;{4\pi\,C\over\alpha^2}\,{T_p\over\sqrt{s}}.
\end{equation}

The striking feature of this model is that, in order to fit
the collider data, the inelasticity must {\em decrease} as
energy increases---from $\sim0.5$ at low energy to $<0.2$ 
for $pp$ interactions in the Fly's Eye energy range \cite{Ding}.
To compensate and obtain a reasonably shallow depth of maximum,
they use a naive (KNP) assumption for the inelasticity
as well as a rapidly increasing cross  section.  Since
both effects increase with energy, the mean inelasticity
is decreased more at high energy.  In other words, assumptions
which reduce the depth of maximum at one energy also tend
to decrease the elongation rate (logarithmic slope of $X_{max}$).

In view of the model dependence
of depth of maximum and the significant systematic
corrections involved in comparing to the data, it is important to
look at the distribution of depth of maximum as well as its mean value.
This is illustrated in Fig. 3 \cite{Gaisseretal}, which shows the contribution
of protons and iron separately to the overall distribution for
the group of events with $E>10^{18}$~eV.  
The calculations were made with the KNP model.
It appears that in this energy
range comparable numbers of light and heavy nuclei are
needed to fit the data.

{\bf Acknowledgments}.  I am grateful to Bob Fletcher, George Frichter,
Paolo Lipari, Todor Stanev and Lou Voyvodic for useful discussions
and collaboration on this work, and I thank Paolo Lipari for
helpful comments on the paper.  This research is supported in
part by the U.S. Department of Energy under Grant No. DE-FG02-91ER40626.

\vspace{1cm}

\noindent
CAPTIONS

\vspace{.5cm}

\noindent
Fig. 1.
Inelasticity vs. Energy: Points show the SIBYLL
interaction model; the lines represent Eq.~\ref{inelasticity} (see text).

\vspace{.5cm}

\noindent
Fig. 2. Mean depth of shower maximum vs. primary energy.  Data points
from Fly's Eye.  See text for explanation of lines.

\vspace{.5cm}

\noindent
Fig. 3.
Comparison of the fitted depth of maximum distribution
to Fly's Eye data for $E>10^{18}$~eV \cite{Gaisseretal}.
Contributions of protons (dashed) and iron (dotted) are shown
separately.


\begin{thebibliography}{999}
\bibitem {KK} N.N. Kalmykov \& G.B. Khristiansen, J. Phys. G 21
(1995) 1279.
\bibitem {Ptuskin} V.S. Ptuskin et al., Astron. Astrophys. 268 (1993) 726.
\bibitem {Biermann} P.L. Biermann, Astron. Astrophys. 271 (1993) 649.
\bibitem {Stanevetal} Todor Stanev, P.L. Biermann, T.K. Gaisser,
Astron. Astrophys. 274 (1993) 902.
\bibitem {Biermannetal} P.L. Biermann, T.K. Gaisser \& Todor Stanev,
Phys. Rev. D51 (1995) 3450.
\bibitem {Peters} B. Peters, Il Nuovo Cimento (Serie X) 22 (1961) 800.
\bibitem {Zatsepin} G.T. Zatsepin, N.N. Gorunov \& L.G. Dedenko,
Izv. Akad. Nauk USSR ser. Fiz. 26 (1962) 685.
\bibitem {Swordy} Simon Swordy, Proc. XXIII Int.
Cosmic Ray Conference (Calgary,  ed. D.A. Leahy, R.B. Hicks
\& D. Venkatesan, World Scientific, 1993) p. 243.
\bibitem {JACEE} K. Asakamori et al. (JACEE Collaboration)
Proc. XXIII Int. Cosmic Ray Conf. (Calgary, 1993) pp. 21 and 25.
\bibitem {LagageCesarsky} P.O. Lagage \& C.J. Cesarsky, Astron. Astrophys.
118 (1983) 223 and 125 (1983) 249.
\bibitem {Bird} D.J. Bird et al., Phys. Rev. Letters 71 (1993) 3401.
\bibitem {Yakutsk} M.N. Dyakonov et al., Proc. XXIII Int. Cosmic
Ray Conf. (Calgary) 4 (1993) 303.
\bibitem {Knapp} J. Knapp, D. Heck, S.S. Ostapchenko \& G. Schatz,
``Comparison of Hadronic Interaction Models used in Air
Shower Simulations and of their Influence on Shower Development
and Observables'' (Draft, July 31, 1996).  See also papers by
J. Knapp and D. Heck in these Proceedings.
\bibitem {cap} A. Capella, U. Sukhatme, C.-I. Tan \& J.
Tran Thanh Van, Physics Reports 236 (1994) 225.
\bibitem {Ranft} J. Ranft, Phys. Rev. D51 (1995) 64 and
G. Battistoni, C. Forti \& J. Ranft, Astroparticle Physics 3 (1995) 157.
\bibitem {QGS} A.B. Kaidalov, K.A. Ter-Martirosyan \& Yu. M. Shabelsky,
Yad. Fiz. 43 (1986) 1282.
\bibitem {QGSjet} N.N. Kalmykov \& S.S. Ostapchenko,
Phys. At. Nucl. 56 (1993) 346.
\bibitem {Werner} K. Werner, Physics Reports 232 (1993) 87.
\bibitem {Durand} L. Durand and H. Pi, Phys. Rev. Lett. 58 (1987) 303;
Phys. Rev. D38 (1988) 78.
\bibitem {GS} T.K. Gaisser \& Todor Stanev, Phys. Letters
219 (1989) 375.
\bibitem {SIBYLL} R.S. Fletcher et al., Phys. Rev. D50 (1994) 5710.
\bibitem {KNP} B.Z. Kopeliovich, N.N. Nikolaev \& I.K. Potashnikova,
Phys. Rev. D39 (1989) 769.
\bibitem {MOCCA} A.M. Hillas, Proc. 24th Int. Cosmic Ray Conf. (Rome)
vol. 1 (1995) 270.
\bibitem {Gaisser} T.K. Gaisser et al., Phys. Rev. D47 (1993) 1919.
\bibitem {HK} J. H\"{u}fner \& A. Klar, Phys. Letters 145B (1984) 167.
\bibitem {Barton} D.S. Barton et al., Phys. Rev. D27 (1983) 2580.
\bibitem {deMarzo} C. de Marzo et al., Phys. Rev. D26 (1982) 1019.
\bibitem {Glauber} R.J. Glauber \& G. Matthiae, Nucl. Phys.
B21 (1970) 135.
\bibitem {CHL} D. Cline, F. Halzen and J. Luthe, Phys. Rev. Lettters 
31 (1973) 491.
\bibitem {GH} T.K. Gaisser and F. Halzen, Phys. Rev. Letters 54 (1987) 1754.
\bibitem {PS} G. Pancheri and Y. Srivastava, Phys. Lett. B 159 (1985) 69.
\bibitem {Sjostrand} H. Bengtsson and T. Sj\"{o}strand,  
Comput. Phys. Commun. 46 (1987) 43.
\bibitem {Frichter} George Frichter et al., (in preparation).
\bibitem {TKG} T.K. Gaisser {\em Cosmic Rays and Particle Physics}
(Cambridge University Press, 1990.
\bibitem {Akeno} N. Hayashida et al., J. Phys. G 21 (1995) 1101.
See also the talk of M. Nagano at this conference.
\bibitem {Serap} Serap Tilav, Ph.D. Thesis, University of Delaware (1991).
\bibitem {Ding} H.Y. Dai et al., Ap.J. (1996) to be published.
\bibitem {Chou-Yang} T.T. Chou and C.N. Yang, Phys. Rev. D32 (1985) 1692.
\bibitem {Gaisseretal} T.K. Gaisser et al., Comments on Astrophysics 17
(1993) 103.
\end{thebibliography}
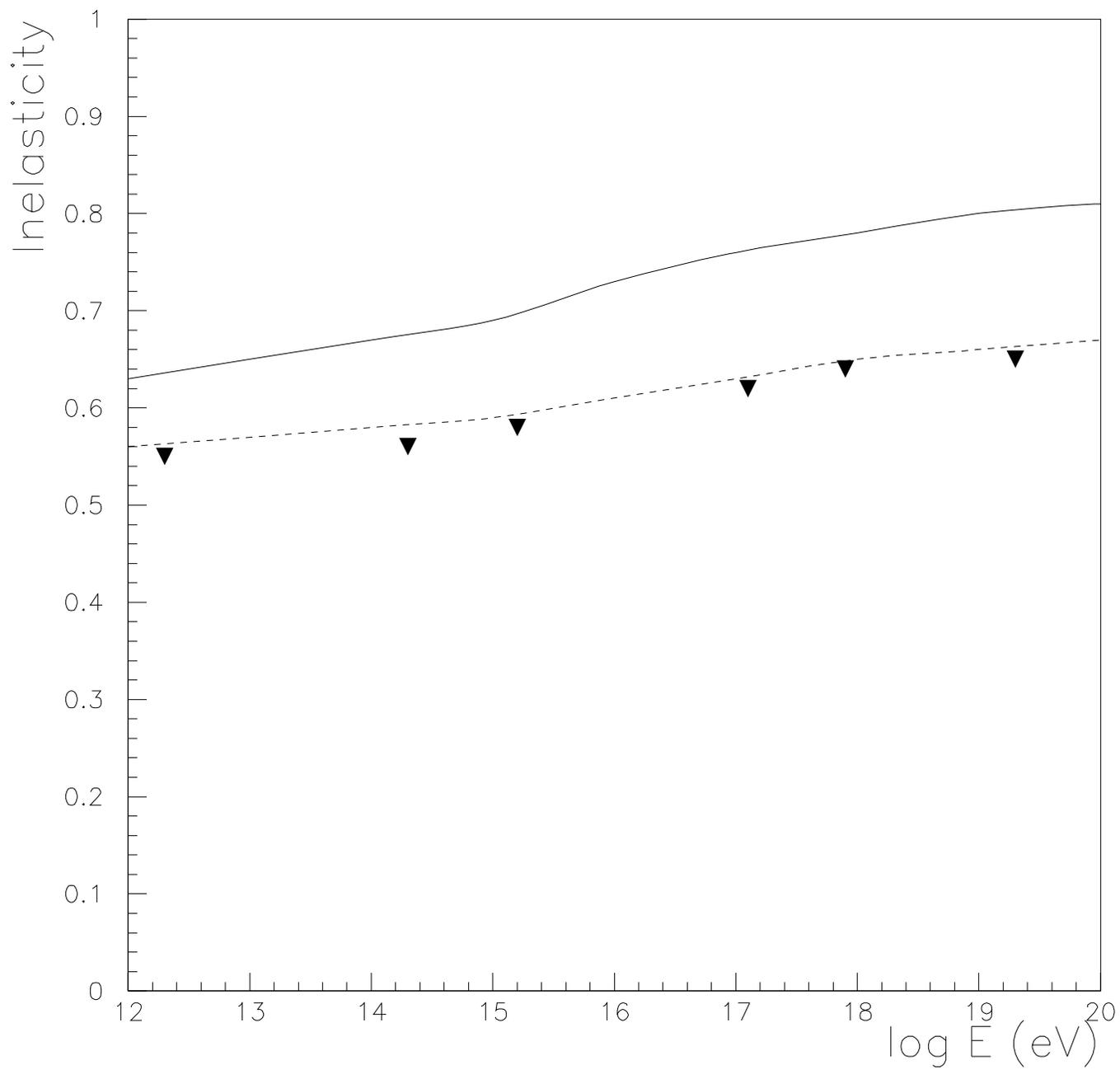
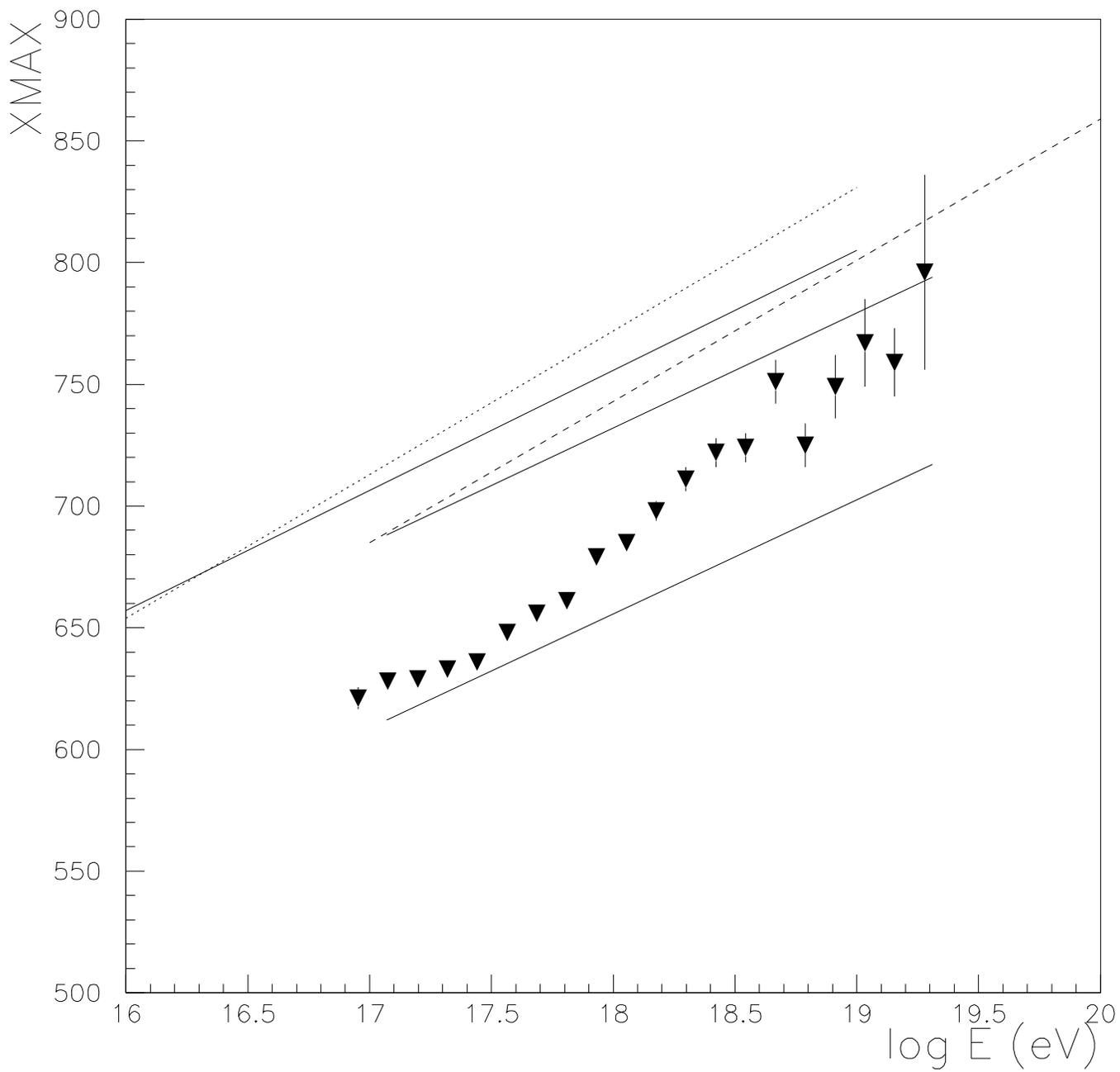
\end{document}